# Gate-tunable heavy fermions in a moiré Kondo lattice

Wenjin Zhao[1], Bowen Shen[2], Zui Tao[2], Zhongdong Han[3], Kaifei Kang[2], Kenji Watanabe[4], Takashi Taniguchi[4], Kin Fai Mak[1,2,3*], Jie Shan[1,2,3*]

[1]Kavli Institute at Cornell for Nanoscale Science, Ithaca, NY, USA
[2]School of Applied and Engineering Physics, Cornell University, Ithaca, NY, USA
[3]Laboratory of Atomic and Solid State Physics, Cornell University, Ithaca, NY, USA
[4]National Institute for Materials Science, 1-1 Namiki, 305-0044 Tsukuba, Japan

Email: kinfai.mak@cornell.edu; jie.shan@cornell.edu

**The Kondo lattice, describing a matrix of local magnetic moments coupled via spin-exchange interactions to itinerant conduction electrons, is a prototype of strongly correlated quantum matter [1–5]. Traditionally, Kondo lattices are realized in intermetallic compounds containing lanthanide or actinide [1,2]. The complex electronic structure and limited tunability of both the electron density and exchange interactions in these bulk materials pose significant challenges to study Kondo lattice physics. Here, we report the realization of a synthetic Kondo lattice in AB-stacked MoTe₂/WSe₂ moiré bilayers, where the MoTe₂ layer is tuned to a Mott insulating state, supporting a triangular moiré lattice of local moments, and the WSe₂ layer is doped with itinerant conduction carriers. We observe heavy fermions with a large Fermi surface below the Kondo temperature. We also observe destruction of the heavy fermions by an external magnetic field with an abrupt decrease of the Fermi surface size and quasiparticle mass. We further demonstrate widely and continuously gate-tunable Kondo temperatures through either the itinerant carrier density or Kondo interaction. Our study opens the possibility of in-situ access to the rich phase diagram of the Kondo lattice with exotic quantum criticalities in a single device based on semiconductor moiré materials [2–4,6–11].**

## Main text

Moiré materials provide a highly tunable platform to explore the strongly correlated electron phenomena [7,8,12,13]. A series of correlated insulating, magnetic, and superconducting states has emerged as a result of the flat electronic bands induced by the moiré superlattices [7,8,12–14]. In particular, a Mott insulator with local magnetic moments has been realized in semiconductor moiré materials at half filling of the moiré bands [15–17]. The strong electronic interactions open a Mott gap with a fully filled Hubbard band. Coupling the itinerant electrons to the lattice of local moments via the exchange interaction has been proposed as a route to realizing a Kondo lattice [18–23].

Here, we experimentally demonstrate the realization of a moiré Kondo lattice in AB-stacked (60-degree-aligned) MoTe₂/WSe₂ bilayers with hole doping. These bilayers form a triangular moiré lattice with period of about 5 nm, corresponding to a moiré density of $n_M \approx 5 \times 10^{12}$ cm⁻², as a result of the 7% mismatch between the MoTe₂ and WSe₂ lattices [24–26]. The Wannier orbitals of the topmost MoTe₂ and WSe₂ valence bands are centered at two distinct high-symmetry stacking sites of the moiré lattice. Together, they form a honeycomb lattice [27–30]. The moiré potential is substantially stronger at the Mo-site than the W-site (see Methods for more discussions). Hence, the first MoTe₂ moiré band is relatively flat and the first WSe$_2$ moiré band is dispersive so that the holes in the MoTe$_2$ layer can be localized by interactions, whereas the holes in the WSe$_2$ layer remain itinerant. This is supported by density functional calculations [27], as well as experimental results discussed below.

In dual-gated devices (Fig. 1a), the two gate voltages can independently tune the total electrostatic doping density, $\nu$ (in units of $n_M$), in the MoTe$_2$/WSe$_2$ bilayer, and the electric field, $E$, perpendicular to the bilayer. The electric field tunes the interlayer potential difference or band alignment [24]. An interesting scenario arises when the first WSe$_2$ moiré band is slightly below the MoTe$_2$ moiré band. The interlayer hybridization, denoted by the interlayer hopping integral $t_\perp$ (Fig. 1b), remains weak because, to a very good approximation, the two bands have opposite spins for the same valley and the interlayer tunneling is nearly spin forbidden [26,27]. The system has been shown to undergo gate-tuned topological phase transitions at both $\nu = 1$ and $\nu = 2$ upon band inversion [24,26]. Particularly, at $\nu = 1$ it is a Mott (or charge transfer) insulator before the band inversion and a quantum anomalous Hall insulator after the band inversion [24,30].

A simple Kondo lattice can be envisioned before the band inversion. It corresponds to the upper shaded region of the $(\nu, E)$ phase diagram in Fig. 1c. The total doping density is $\nu = 1 + x$, the hole density in the MoTe$_2$ layer $\nu_{Mo}$ is fixed at 1, and the hole density in the WSe$_2$ layer $\nu_W$ is $x > 0$. This is achievable for a range of electric field, for which the dispersive WSe$_2$ Mott band is inside the MoTe$_2$ Mott gap, and the Fermi level cuts through the WSe$_2$ band [18,19] (insets in Fig. 1c). The electric-field span of the region, $\Delta E$, provides a measure of the Mott gap, $U = d\,\Delta E$, where $d$ is the interlayer dipole moment of the bilayer [31,32]. Furthermore, since the Fermi energy scales linearly with density for massive electrons in two dimensions, the boundaries of the Kondo lattice region shift linearly with $x$.

The possibility of separating the local moments and conduction holes into two different layers provides remarkable gate tunability of the Kondo effect [18,19,21]. The Kondo coupling, $J_K$, between a conduction hole and local moment can be estimated $\sim 2t_\perp^2(\frac{1}{\Delta} + \frac{1}{U-\Delta})$ by considering an interlayer (antiferromagnetic) super-exchange interaction, where $\Delta$ is the charge transfer gap [18,19]. Bringing the WSe$_2$ band close to resonance with either Hubbard band by the electric field can significantly enhance $J_K$. When the Kondo coupling effects dominate, the local moments hybridize with the conduction holes to produce a large Fermi surface with heavy quasiparticle masses. The onset temperature for Kondo screening increases with $J_K$ and the conduction hole density.

## A moiré Kondo lattice

We perform magneto transport and optical spectroscopy measurements to identify the discussed Kondo lattice region. Details on the device fabrication and measurements are provided in Methods. Figure 2a shows the longitudinal resistance, $R_{xx}$, as a function of $(\nu, E)$ under an out-of-plane magnetic field $B = 13.6$ T at temperature $T = 1.6$ K. The doping density and electric field are determined from two gate voltages using the parallel plate capacitor model, and the density is independently calibrated by quantum oscillation measurements. The grey regions cannot be accessed in the current device structure either due to the large sample/contact resistance or limited accessible gate voltages. Quantum oscillations (stripes) are observable in many regions of the phase diagram, indicating Fermi liquid behavior.

For $\nu = 1 + x$, we identify a region with pronounced vertical stripes (region II). Inside, the Landau levels do not disperse with $E$, that is, the electric field does not affect the Fermi surface. This is consistent with the Fermi level located inside the Mott gap with $\nu_{Mo} = 1$ and $\nu_W = x$. The level degeneracy is also consistent with the formation of spin-valley-polarized Landau levels only in the WSe$_2$ layer. In contrast, in the region right below II with smaller electric fields, the vertical stripes turn into diagonal ones, and a second set of weaker stripes emerges and intersects the first set. The prominent quantum oscillations arise from the WSe$_2$ layer, and the weaker ones, from the MoTe$_2$ layer, because the latter has heavier mass and lower mobility. The Fermi level cuts through bands of both layers. The stripes disperse with $E$ because the field varies the relative alignment of the Landau levels from two different layers. We estimate the Mott gap of region II to be about 32 meV from the electric-field span of the region, assuming the interlayer dipole moment to be $d \approx e \times 0.26$ nm (from optical measurements, $e$ denoting the elementary charge) [25].

Pronounced vertical stripes can also be identified in two other regions: region I with $\nu = x$ and region III with $\nu = 2 + x$. Similarly, they correspond to quantum oscillations in the WSe$_2$ layer with density $x$; the Fermi level is located inside an energy gap of the MoTe$_2$ layer, particularly, the large semiconductor band gap in region I ($\nu_{Mo} = 0$) and the gap between the first and second moiré bands in region III ($\nu_{Mo} = 2$). The full extent of region I is beyond the accessible gate voltages. Assuming the same interlayer dipole moment $d$, we estimate the moiré band gap from the electric-field span of region III to be about 16 meV. Both the Mott and moiré band gaps depend on $B$ due to the Zeeman effect (Fig. 2b and Extended Data Fig. 1, also see Methods). The zero-field Mott and moiré band gaps (20 and 26 meV, respectively) are inferred from the layer-resolved exciton optical response (Extended Data Fig. 2).

We can also determine the quasiparticle mass of WSe$_2$ from the temperature dependence of the quantum oscillations (Fig. 2c and Extended Data Fig. 3). At 13.6 T, the value is around $0.5\,m_0$ in all three regions ($m_0$ denoting the electron mass). The value is slightly larger than the hole mass in monolayer WSe$_2$ ($\approx 0.45\,m_0$) [33]. This result supports a dispersive WSe$_2$ band and weak moiré potential at the W-sites.

The simultaneous existence of a sizeable Mott gap in MoTe$_2$, which supports a triangular lattice of local moments, and of itinerant holes in WSe$_2$ in region II provides the key ingredients for a Kondo lattice. Below we probe whether a heavy-fermion liquid exits. Since the Kondo effect is not expected in region III, where the first MoTe$_2$ moiré band is fully filled with two holes of opposite spin per moiré unit cell, we use region III as a control experiment (Fig. 1). The full ($\nu, E$) phase diagram is schematically illustrated in Extended Data Fig. 4.

**Emergence of heavy fermions**

Figure 3a and 3b are the temperature dependence of the resistance under zero magnetic field in region II ($E = 0.645$ V/nm) and III ($E = 0.45$ V/nm), respectively. Different curves correspond to different doping densities at a fixed $E$, and the same color denotes the same hole density in the WSe$_2$ layer. In both regions, $R_{xx}$ decreases with increasing $x$, and exhibits a $T^2$-dependence at low temperatures (insets). (There is also a small increase in $R_{xx}$ below 10 K for $x \lesssim 0.15$ in region III; its origin is not understood at this point.) In region II, there is a characteristic temperature $T^*$, below which resistance drops significantly and whose value increases with $x$. No resistance peak or onset of significant resistance drop is observed in region III.

The $T^2$-dependence of $R_{xx}$ at low temperatures is a characteristic of a Landau Fermi liquid. We fit the dependence using $R_{xx} = R_0 + AT^2$, where parameter $R_0$ is the residual resistance limited by impurity scattering, and $A^{1/2}$ is linearly proportional to the quasiparticle mass in the Fermi liquid theory [34]. The doping dependence of $A^{1/2}$ is shown in Fig. 3c. The value in region II is more than an order of magnitude larger than in region III. The large mass enhancement points to the emergence of heavy fermions in region II.

We probe the size of the Fermi surface by measuring the Hall resistance, $R_{xy}$. Figure 3d shows the magnetic-field dependence of the Hall coefficient, $R_H = R_{xy}/B$, for representative doping densities in region II and III. The corresponding result for $R_{xx}$ is included in Extended Data Fig. 5. In region III, $R_H$ weakly depends on the field up to 14 T. Oscillatory features appear above $\sim 4$ T from the formation of Landau levels. Conversely in region II, $R_H$ has an opposite sign at low fields; it exhibits an abrupt change including a sign switch around a critical field, $B_c \sim 6$ T; and above $B_c$, it reaches a similar value as in region III for the same $x$. The oscillatory features are not observable below $B_c$ and become visible immediately above $B_c$.

For a given doping density, the magnetoresistance at varying temperatures follows the Kohler's scaling (Extended Data Fig. 6). We therefore assume a single charge carrier type and evaluate the Hall density, $\nu^*$ (in units of $n_M$), from the Hall coefficient as $1/(eR_H n_M)$ for $x = 0.35$ (Fig. 3e). In region III, we obtain $\nu^* \approx -x$, which indicates a hole Fermi surface with density $x$. This is expected because the first MoTe$_2$ moiré band is fully filled with 2 holes, and the Fermi level cuts through the WSe$_2$ moiré band inside the MoTe$_2$ moiré band gap. The Hall density is also consistent with the value extracted from quantum oscillations. In region II, below $B_c$ we obtain $\nu^* \approx 1 - x$. It indicates an electron Fermi surface with density $1 - x$, which is equivalent to a hole density of $1 + x$ because the band degeneracy is 2. This result shows that the local moments in the MoTe$_2$ layer are hybridized with the conduction holes in the WSe$_2$ layer to form a large hole Fermi surface. The observed large Fermi surface, in combination with the quasiparticle mass enhancement, supports the realization of a Kondo lattice in region II [35]. We use the peak or characteristic temperature $T^*$, below which $R_{xx}$ decreases significantly, to measure the Kondo temperature. The Kondo temperature is much smaller than the itinerant hole Fermi temperature.

**Magnetic destruction of heavy fermions**

For $B > B_c$, the Hall measurement in region II shows that the large Fermi surface with hole density $1 + x$ is reduced to a small Fermi surface with hole density $x$. The Fermi surface size reduction is also correlated with the emergence of a 'light-fermion liquid', manifested in the immediate appearance of quantum oscillations. We determine the quasiparticle mass from the temperature dependence of the quantum oscillations to be $\sim 0.5\ m_0$, which is comparable to the value in region I and III (Fig. 2c).

Magnetic destruction of heavy fermions is possible when the Zeeman energy exceeds the Kondo temperature, below which the Kondo singlets emerge [35–38]. We estimate the critical Zeeman energy ($g\mu_B B_c \sim 3$ meV at $x \approx 0.3$) using the hole g-factor, $g \approx 10$, in monolayer transition metal dichalcogenides [39,40] ($\mu_B$ is the Bohr magneton). It is in good agreement with the corresponding Kondo temperature ($T^* \approx 40$ K). Magnetic destruction of heavy fermions could be a metamagnetic phase transition or a smooth crossover [35,41]. The significant narrowing of the $R_H$ sign change region with decreasing temperature (Fig. 3f) suggests a metamagnetic phase transition [42]. Metamagnetic transitions from a Kondo-singlet paramagnet to a spin-polarized paramagnet have been reported in rare-earth heavy fermion compounds [36,37]. Future experiments that correlate the Hall and magnetization measurements are required to understand the nature of the transition in moiré Kondo lattices.

**Gate-tunable coherent Kondo effect**

Finally, we demonstrate continuous gate tuning of the Kondo effect. We study the temperature dependence of the resistance and extract the Kondo temperature $T^*$ and the parameter $A^{-1/2}$ in region II (Fig. 4 and Extended Data Fig. 7). We show one cut along $E = 0.645$ V/nm (Fig. 4a) and another cut along $x = 0.23$ (Fig. 4b). The Kondo temperature (top panels) can be widely tuned by both doping and electric field. And $A^{-1/2}$ (lower panels), the inverse of the mass enhancement, largely follows $T^*$ as expected [43]. Increasing the doping density is expected to strengthen the Kondo effect since there are more conduction holes to screen the local moments [18,19,43]; this is consistent with the observed dependence of $T^*$ on $x$. However, we observe a stronger enhancement of the Kondo effect when the WSe$_2$ band is close to the MoTe$_2$ lower Hubbard band ($E \approx 0.66$ V/nm) than the upper Hubbard band ($E \approx 0.6$ V/nm). This seemingly disagreement with the simple estimate of the Kondo coupling based on the superexchange interaction [18] reflects the asymmetry in the two Hubbard bands (Methods) as well as potentially other interaction mechanisms that are required in future models to describe our experiment.

In conclusions, we have realized a Kondo lattice in AB-stacked MoTe$_2$/WSe$_2$ moiré bilayers when the WSe$_2$ layer is doped with itinerant holes and the MoTe$_2$ layer remains in a Mott insulating state. We have observed a heavy-fermion liquid with a large Fermi surface. Our work demonstrates a highly gate-tunable Kondo effect. It opens exciting opportunities to study the gate-controlled quantum phase transitions by extending the Kondo temperature further down to zero, for instance, by reducing the doping density in higher-quality devices or the interaction effect in small-twist-angle bilayers [18].

## Methods
### Device fabrication
We fabricated dual-gated MoTe$_2$/WSe$_2$ devices using the layer-by-layer dry transfer technique as previously reported [24,25,44]. We aligned the MoTe$_2$ and WSe$_2$ monolayers by first determining the crystallographic orientations of each monolayer using the optical second-harmonic generation technique [15,16]. The 0- and 60-degree-aligned bilayers were further sorted according to their distinct electric-field dependence of the longitudinal resistance at $\nu = 1$ (Ref. [24,25]). MoTe$_2$ is air sensitive and was handled in a glovebox with O$_2$ and H$_2$O concentration below 1 part per million.

We first fabricated the bottom gates, consisting of hexagonal boron nitride (hBN) of $\sim$ 10-nm thickness and few-layer graphite on silicon substrates with pre-patterned Ti/Au electrodes. The polycarbonate residue from the dry transfer process was removed in chloroform for 2 hours. We deposited 5-nm Pt contacts on hBN by standard electron-beam lithography and evaporation, followed by another step of electron beam lithography and metallization to form 5-nm Ti/40-nm Au to connect the thin Pt contacts on hBN to pre-patterned electrodes. We cleaned the device surface after lift-off using the atomic force microscope (AFM) in the contact mode. For the top gates, we chose a relatively thin hBN layer ($\sim$ 4 nm) to achieve high breakdown electric fields on the order of 1 V/nm. We also chose a narrow top graphite gate electrode, which defines the region of interest in the Hall bar geometry. The results in the main text were obtained from device 1. All results were reproduced in device 2 (Extended Data Fig. 8).

### Electrical measurements
Electrical measurements were performed in a closed-cycle $^4$He cryostat with a $^3$He insert (Oxford TeslatronPT) with magnetic field up to 14 T and temperature down to 300 mK. The standard low-frequency (10 - 20 Hz) lock-in technique was used to measure the sample resistance at low bias (0.2 - 1 mV). A voltage pre-amplifier with 100-M$\Omega$ impedance was used to measure the sample resistance up to about 10 M$\Omega$. The longitudinal and transverse voltage drops and the source-drain current were recorded. Finite longitudinal-transverse coupling occurs in our devices. We used the standard procedure to obtain the longitudinal and Hall resistance by symmetrizing $\frac{R_{xx}(B) + R_{xx}(-B)}{2}$ and anti-symmetrizing $\frac{R_{xy}(B) - R_{xy}(-B)}{2}$ the measured $R_{xx}$ and $R_{xy}$ values under positive and negative magnetic fields, respectively.

### Estimate of the bandwidths
The density functional theory calculations and our experiment show that the moiré potential is substantially stronger for holes in the MoTe$_2$ layer than the WSe$_2$ layer [27]. A plausible explanation is that lattice corrugations and strain-induced band energy shift contribute substantially to the moiré potential in AB-stacked MoTe$_2$/WSe$_2$ bilayers; the Young's modulus of WSe$_2$ is about 1.7 times of MoTe$_2$, and lattice corrugations under lattice reconstruction in WSe$_2$ are much smaller [45,46].

We estimate the upper limit of the width of the first WSe$_2$ moiré band from the doping density and density of states. Assuming a parabolic band of mass $m_W$, the density of states is $\frac{dn}{dE} = \frac{m_W}{\pi \hbar^2}$, where $\hbar$ is the reduced Planck constant. The total density of the first moiré band is $2n_M \approx 1 \times 10^{13}$ cm$^{-2}$. We obtain the upper bound for the dispersive WSe$_2$ bandwidth to be about 48 meV using $m_W \approx 0.5\, m_0$ from experiment.

The MoTe$_2$ lower Hubbard bandwidth can be estimated by $d\,\Delta E$, where $d \sim 0.26$ e × nm is the interlayer dipole moment [25], and $\Delta E$ is the electric field difference when the WSe$_2$ valence band edge is aligned with the maximum and minimum of the lower Hubbard band. These two fields are $E \approx 0.7$ V/nm (corresponding to the band inversion point with $\nu_W = \nu_{Mo} = 0$) and $E \approx 0.65$ V/nm (with $\nu_W = 0$ and $\nu_{Mo} = 1$) as shown in Fig. 2a and Extended Data Fig. 2. We estimate the lower Hubbard bandwidth to be 13 meV. Similarly, we estimate the upper Hubbard bandwidth to be 26 meV.

**Determination of the Mott and moiré band gaps**
In the main text, we describe how the Mott and moiré band gaps were determined from the ($\nu$, $E$) map of $R_{xx}$ at 13.6 T from the electric-field span of region II and III, respectively. We performed the same measurements under varying magnetic fields. Extended Data Fig. 1 shows the result under several representative magnetic fields. The quantum oscillations are no longer discernable for $B < 6$ T in region II due to the formation of the heavy fermions, but persist down to 4 T in region III. Figure 2c summarizes the magnetic-field-dependent Mott and moiré band gaps. The zero-field values were obtained from the layer-resolved exciton optical response (Extended Data Fig. 2), which is sensitive to charge doping. In region III, the magnetic field lifts the spin degeneracy of the moiré bands and continuously reduces the moiré band gap by the Zeeman effect. On the other hand, in region II the field first polarizes the local moments, and increases the Mott gap above saturation by introducing an additional Zeeman splitting between the Hubbard bands [15].

**Estimate of the heavy fermion mass**
The magnetic destruction of the Kondo singlets at rather small critical magnetic fields ($\sim$ 6 T) prevents us from directly determining the quasiparticle mass of the heavy fermions from the temperature-dependent quantum oscillations. Quantum oscillation cannot be observed below 6 T for the heavy fermions because of the large quasiparticle mass and the presence of disorders. We therefore can only estimate the heavy fermion mass by assuming that the ratio of the quasiparticle mass in region II and III equals to the ratio of $A^{1/2}$ based on the Kadowaki-Woods scaling [34]. Given the measured WSe$_2$ quasiparticle mass $m_W \approx 0.5\, m_0$ and the ratio of $A^{1/2}$ (10 – 20 for varying $x$) (Fig. 2c and Fig. 3c), the heavy fermion mass is estimated to be 5 - 10 $m_0$.

**Determination of the Kondo temperature**
The Kondo temperature $T^*$ is a crossover temperature scale, below which coherent charge transport in the lattice of Kondo singlets develops [2]. The sample resistance significantly drops below $T^*$. In our experiment, we observe a peak/bump in the temperature dependent resistance (Fig. 3a and Extended Data Fig. 7). Because of the broad feature for a crossover, the numerical accuracy for $T^*$ is limited. The crossover temperature $T^*$ was extracted as the minimum of $|dR_{xx}/dT|$, which corresponds to the resistance peak temperature or the temperature below which the resistance decreases significantly if a peak is absent.

**Acknowledgement**
We thank Liang Fu, Daniele Guerci, Andrew Millis, Antoine Georges, Angel Rubio, Senthil Todadri, Ajesh Kumar, Andrew Potter, Debanjan Chowdhury and Yahui Zhang for fruitful discussions.

**Author contributions**
W.Z. and B.S fabricated the devices. W.Z. and B.S. performed the electrical transport measurements and analyzed the data with the help of Z.H. and K.K. Z.T. and W.Z. performed the optical measurements. K.W. and T.T. grew the bulk hBN crystals. W.Z., J.S. and K.F.M. designed the scientific objectives and oversaw the project. All authors discussed the results and commented on the manuscript.

**Figures**

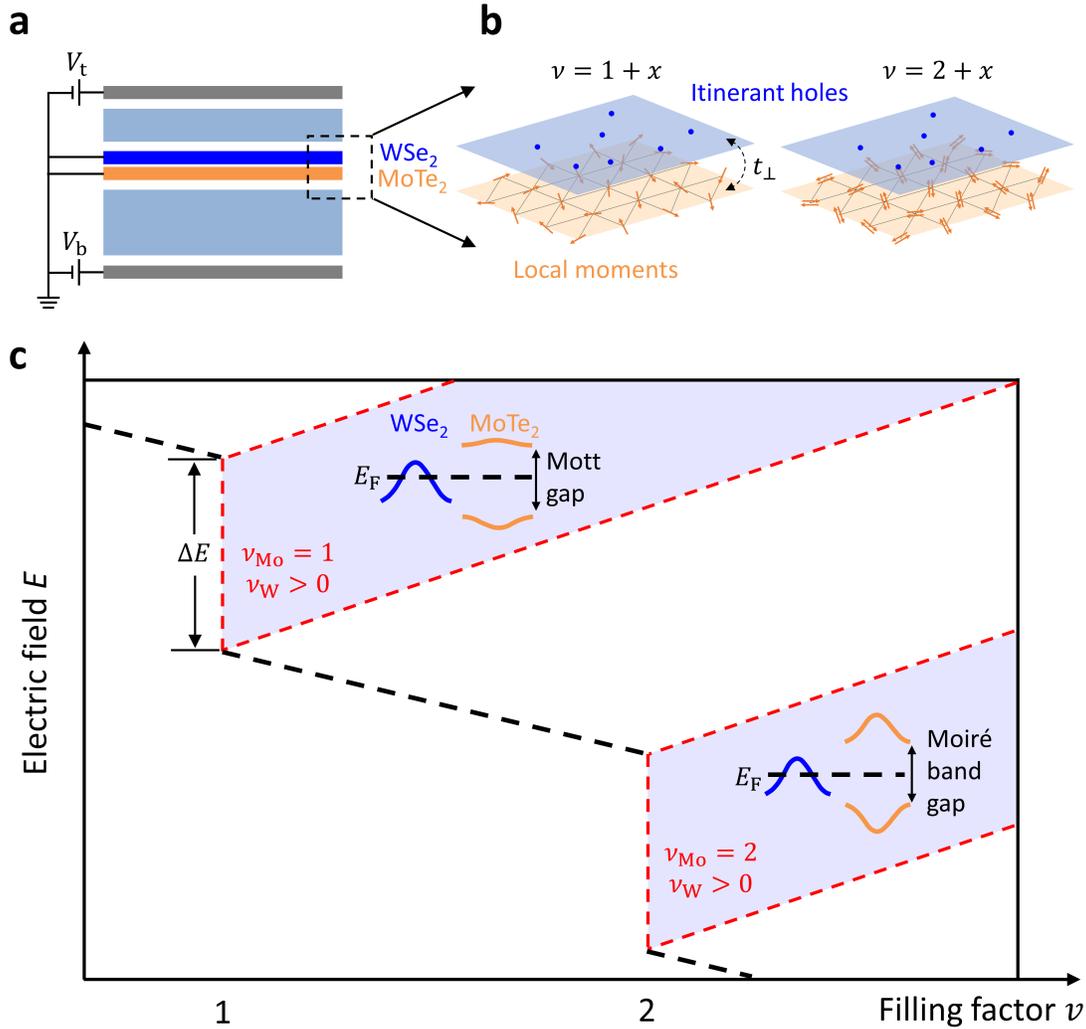

**Figure 1 | Moiré Kondo lattice in AB-stacked MoTe₂/WSe₂. a,** Schematic side view of a dual-gated Hall bar device of AB-stacked MoTe₂/WSe₂. **b,** Schematic illustration of a moiré Kondo lattice (left). The MoTe₂ layer is filled with one hole per moiré site ($\nu_{Mo} = 1$) and the WSe₂ layer hosts the itinerant holes ($\nu_W = x$). Here $t_\perp$ denotes the interlayer hoping. The right schematic shows the corresponding scenario when the MoTe₂ layer is filled with two holes per moiré site ($\nu_{Mo} = 2$) with anti-aligned spins; no Kondo physics is expected in this case. **c,** The $(\nu, E)$ phase diagram with regions defined by different fillings in MoTe₂ and WSe₂. Kondo lattice physics is realized in region II ($\nu_{Mo} = 1$ and $\nu_W = x$); region III ($\nu_{Mo} = 2$ and $\nu_W = x$) provides a control experiment. Insets: the moiré band structure corresponding to region II and III. The Fermi level in region II (III) is located in between the MoTe₂ Hubbard (moiré) bands and goes through the WSe₂ moiré band. The electric field span $\Delta E$ (arrow) is directly proportional to the Mott gap in region II or the moiré band gap in region III. The WSe₂ layer is charge-neutral below the black dashed lines.

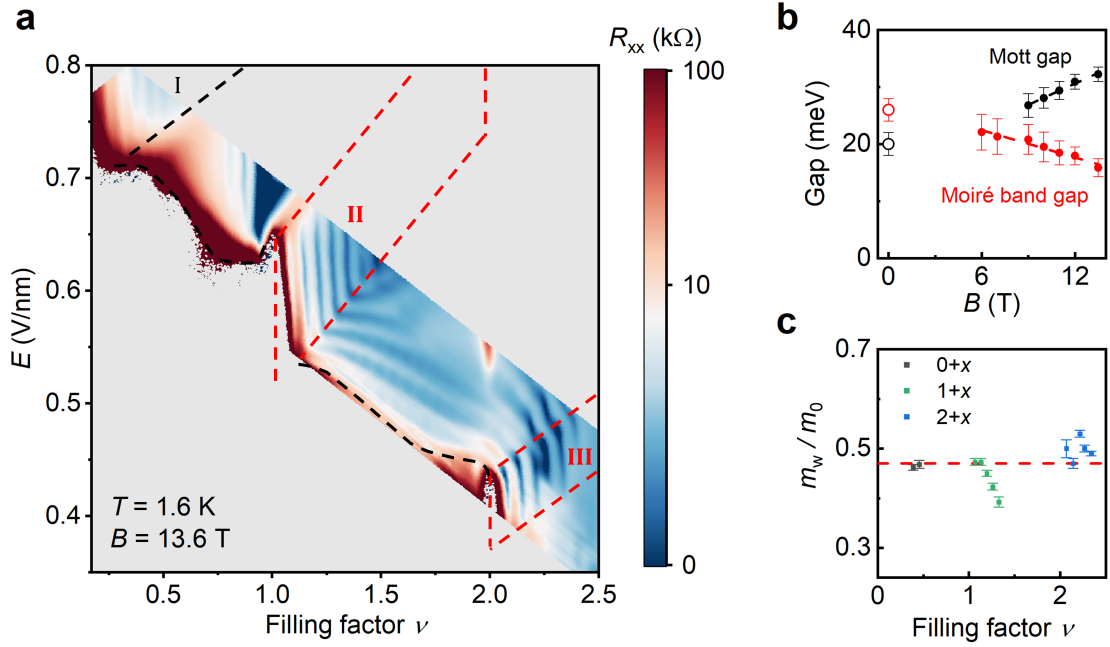

**Figure 2 | Electrostatics phase diagram. a,** Dependence of $R_{xx}$ on the total filling factor $\nu$ and the out-of-plane electric field $E$ at $B = 13.6$ T and $T = 1.6$ K. The dashed lines mark the phase boundaries between the different regions in the electrostatics phase diagram corresponding to Fig. 1c and Extended Data Fig. 4. The red dashed lines mark the boundaries for region II and III. **b,** Magnetic field dependence of the Mott gap and the moiré band gap extracted from the electric field span in region II and III, respectively. The zero magnetic field data (empty dots) are obtained from optical measurements (Extended Data Fig. 2). The dashed lines are guides to the eye. **c,** Filling factor dependence of the $WSe_2$ itinerant hole mass extracted from the temperature dependent quantum oscillations (Extended Data Fig. 3) in region I ($\nu = \nu_{Mo} + \nu_W = 0 + x$), II ($\nu = 1 + x$) and III ($\nu = 2 + x$). Different colored data points correspond to different regions. The dashed line denotes the average hole mass.

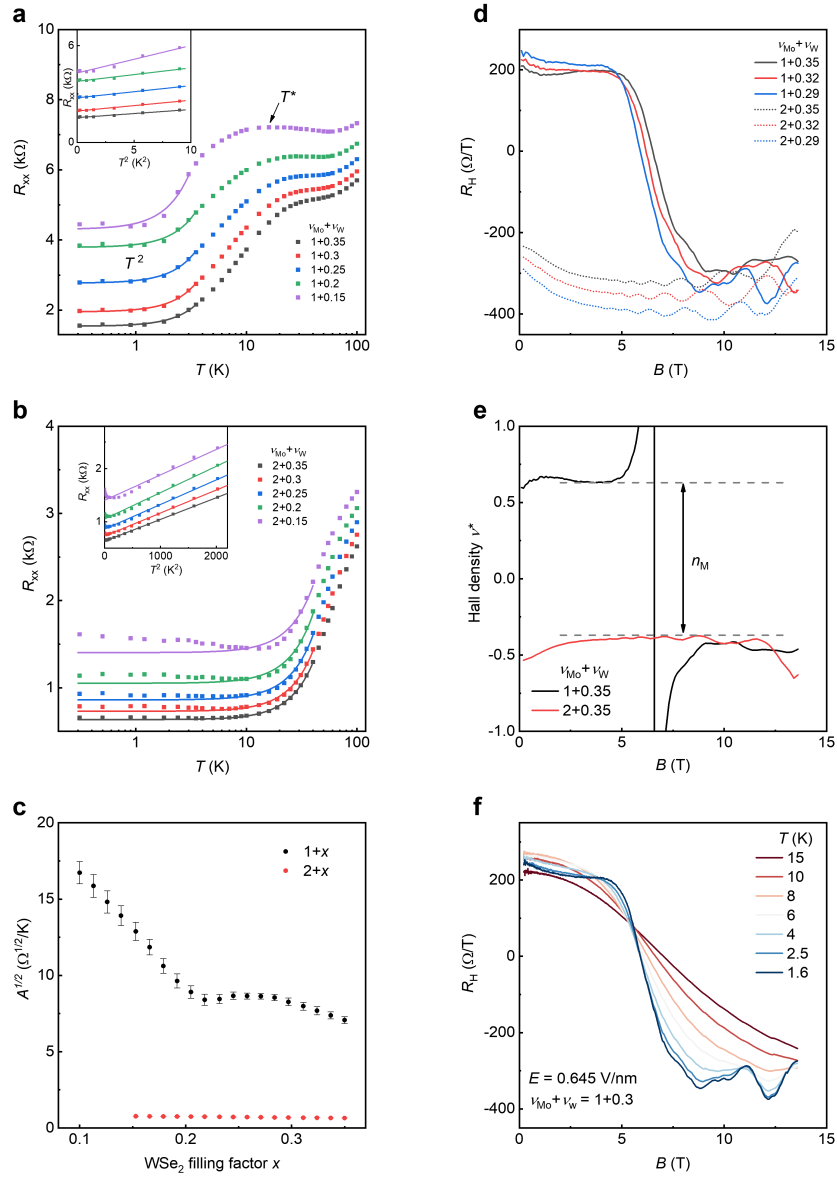

**Figure 3 | Emergence of heavy fermions and magnetic destruction of Kondo singlets.**
**a, b**, Temperature dependent $R_{xx}$ at varying doping densities for $\nu = 1 + x$ (**a**) and $\nu = 2 + x$ (**b**). The solid lines are the best fits to the quadratic temperature dependence at low temperatures. The insets show the scaling of $R_{xx}$ with $T^2$. **c**, Extracted doping dependence of the coefficient $A^{1/2}$ for both $\nu = 1 + x$ (black) and $\nu = 2 + x$ (red). **d**, Magnetic field dependence of the Hall coefficient $R_{H}$ at $T = 1.6$ K for both $\nu = 1 + x$ (solid curves) and $\nu = 2 + x$ (dotted curves). A sharp change in $R_{H}$ at the critical magnetic field is observed only for $\nu = 1 + x$; that at $\nu = 2 + x$ is weakly field dependent. **e**, The corresponding magnetic field dependence of the Hall density $\nu^*$ for $\nu = 1 + 0.35$ and $\nu = 2 + 0.35$. The arrow indicates a change in the Hall density by the moiré density $n_{M}$. **f**, Magnetic field dependence of $R_{H}$ at $\nu = 1 + 0.3$ and varying temperatures.

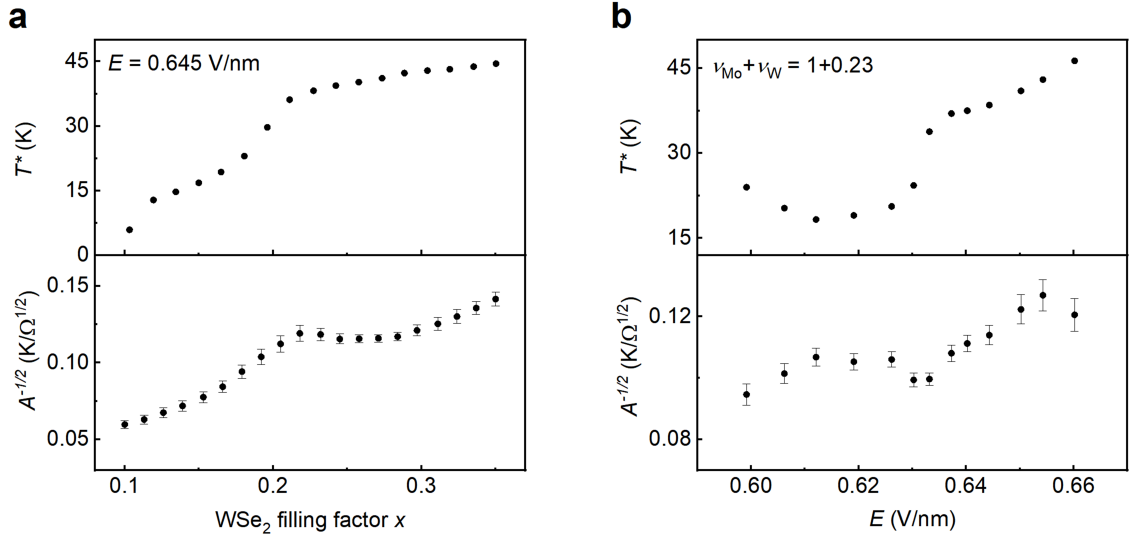

**Figure 4 | Gate-tunable moiré Kondo physics. a**, Extracted doping dependence of $T^*$ (top) and $A^{-1/2}$ (bottom) at a fixed electric field ($E = 0.645$ V/nm) for $\nu = 1 + x$. The two quantities show similar dependences. **b**, Extracted electric field dependence of $T^*$ (top) and $A^{-1/2}$ (bottom) at a fixed filling factor ($\nu = 1 + 0.23$). An asymmetric dependence with respect to the $MoTe_2$ lower ($E \approx 0.66$ V/nm) and upper ($E \approx 0.6$ V/nm) Hubbard bands is observed.

**Extended Data Figures**

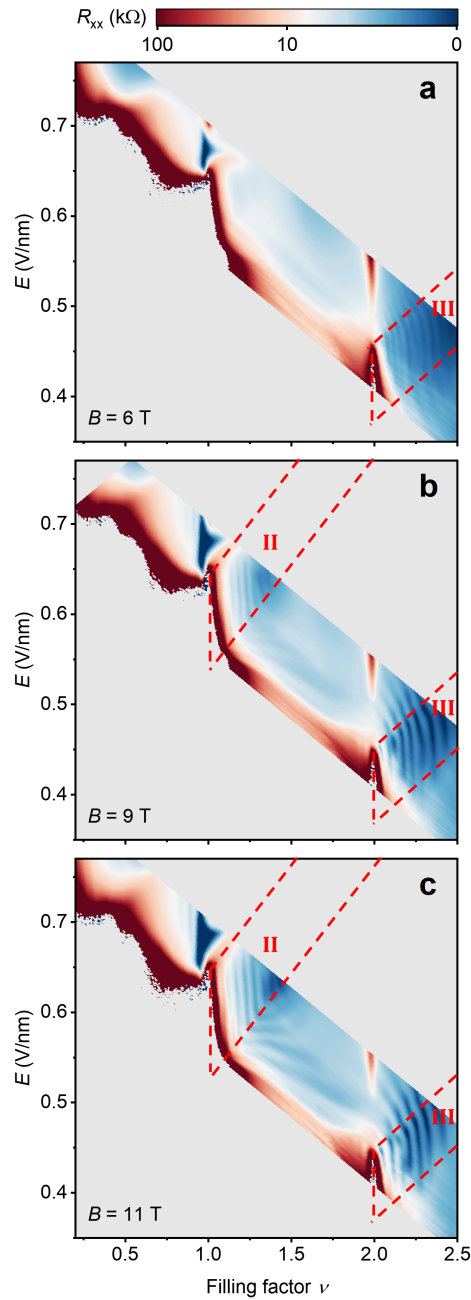

**Extended Data Figure 1 | Filling factor and electric field dependent $R_{xx}$ at different magnetic fields. a, b, c,** Dependence of $R_{xx}$ on the total filling factor $\nu$ and the out-of-plane electric field $E$ at $T = 1.6$ K and $B = 6$ T (**a**), 9 T (**b**) and 11 T (**c**). The dashed lines mark the phase boundaries for region II and III. Region II cannot be identified at $B = 6$ T without Landau levels. The electric field span expands (shrinks) for region II (III) with increasing $B$.

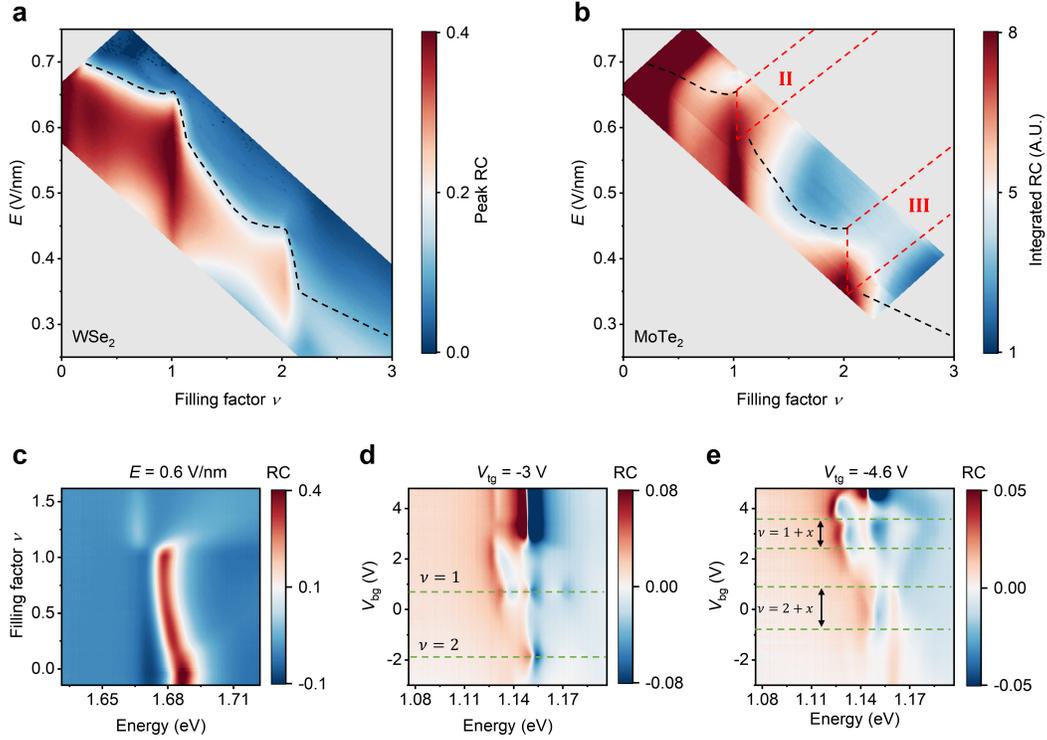

**Extended Data Figure 2 | Filling factor and electric field dependence of exciton reflection contrast. a, b,** The reflection contrast (RC) for the intralayer exciton resonance of WSe₂ (**a**) and MoTe₂ (**b**) as a function of filling factor and electric field at $B = 0$ T and $T = 1.6$ K. The peak RC and the spectrally integrated RC (over the exciton resonance) are shown in **a** and **b**, respectively. **c,** Filling factor dependence of the RC spectrum near the WSe₂ exciton resonance at $E = 0.6$ V/nm. The much weakened neutral exciton resonance accompanied by the appearance of the Fermi polaron resonances for $\nu > 1$ shows that the WSe₂ layer is hole-doped above $\nu = 1$. Repeated measurements at varying electric fields construct the full map in **a**. The sharp drop in the exciton RC with doping helps construct the black dashed line in **a**, above which WSe₂ is hole-doped. **d, e,** Bottom gate voltage ($V_{bg}$) dependence of the RC spectrum near the MoTe₂ exciton resonance at top gate voltages $V_{tg} = -3$ V (**d**) and $V_{tg} = -4.6$ V (**e**). While holes are only doped into the MoTe₂ layer in **d**, holes are shared between the two TMD layers in **e**. Similar to WSe₂, the neutral exciton resonance (near 1.14 eV) is weakened substantially when MoTe₂ is hole-doped (e.g. near $V_{bg} = 3$ V in **d**). The charged exciton resonance is enhanced at the insulating states at $\nu = 1$ and 2 (e.g. green dashed lines in **d**). Multiple moiré exciton resonances are also observed near $\nu = 1$. Similar results are also observed in **e** except now the WSe₂ layer also becomes hole-doped. As a result, there is an extended span in $V_{bg}$ (bound by the green dashed lines and indicated by the arrows) that the MoTe₂ layer is kept in an insulating state (Mott insulator for $\nu = 1 + x$ and moiré band insulator for $\nu = 2 + x$). Repeating the measurements in **d** and **e** at varying $V_{tg}$ constructs the full map in **b**. The extended regions bound by the green dashed lines in **e** together with the black dashed line in **a** help identify region II and III in **b**.

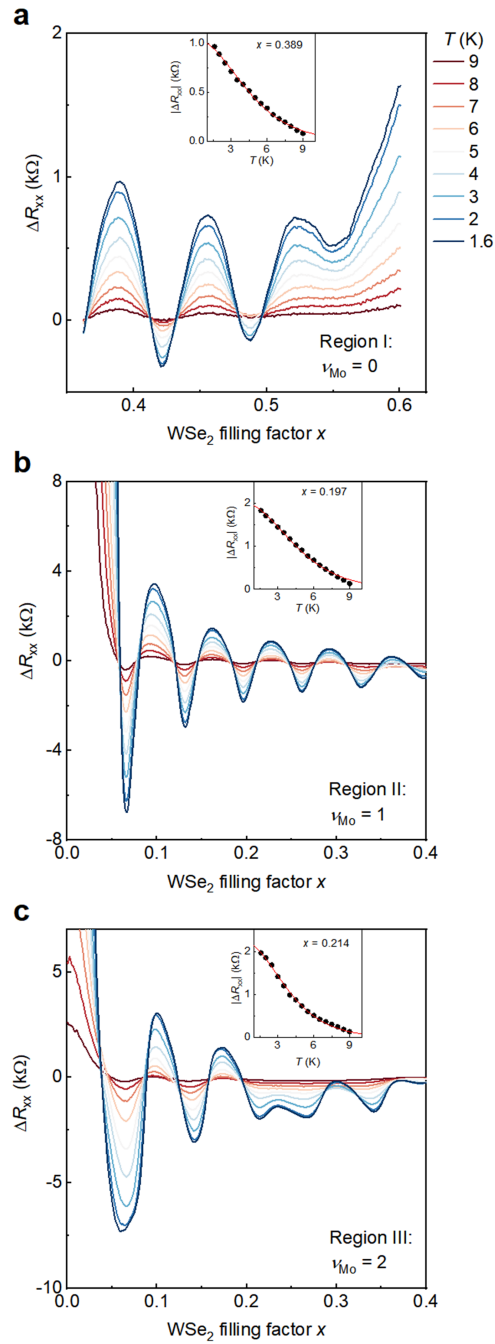

**Extended Data Figure 3 | Determination of the WSe₂ hole mass from quantum oscillations. a**, **b**, **c**, Temperature dependence of the quantum oscillation amplitude $\Delta R_{xx}$ as a function of $\nu_W$ at $B = 13.6$ T for region I (**a**), II (**b**), III (**c**). The filling factor dependence of $R_{xx}$ at $T = 10$ K, where there is no quantum oscillation, is used as a background to obtain $\Delta R_{xx}$. The insets show the fits of $|\Delta R_{xx}|$ versus $T$ to $\Delta R_{xx} = R_a \frac{\lambda(T)}{\sinh \lambda(T)}$ in order to obtain the mass $m_W$. Here $R_a$ is the amplitude and the thermal factor is $\lambda(T) = 2\pi^2 k_B T m_W / \hbar e B$ ($k_B$ denotes the Boltzmann constant).

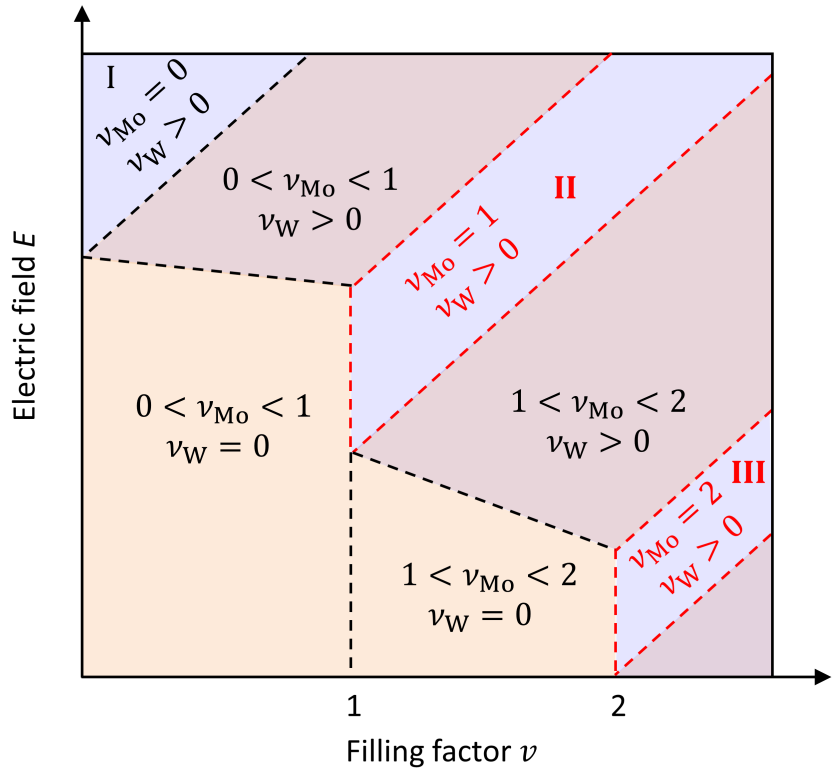

**Extended Data Figure 4 | Schematic electrostatics phase diagram.** The $(\nu, E)$ phase diagram with regions defined by different fillings in $MoTe_2$ and $WSe_2$. Kondo lattice physics is realized in region II ($\nu_{Mo} = 1$ and $\nu_W = x$); region III ($\nu_{Mo} = 2$ and $\nu_W = x$) provides a control experiment. In addition to regions I, II and III discussed in the main text, we can also identify regions with $\nu_W = 0$ and $0 < \nu_{Mo} < 2$, where $WSe_2$ is charge-neutral and only $MoTe_2$ is hole-doped, as well as regions where holes are shared between the two TMD layers ($\nu_W > 0$ and $0 < \nu_{Mo} < 1$ and $\nu_W > 0$ and $1 < \nu_{Mo} < 2$).

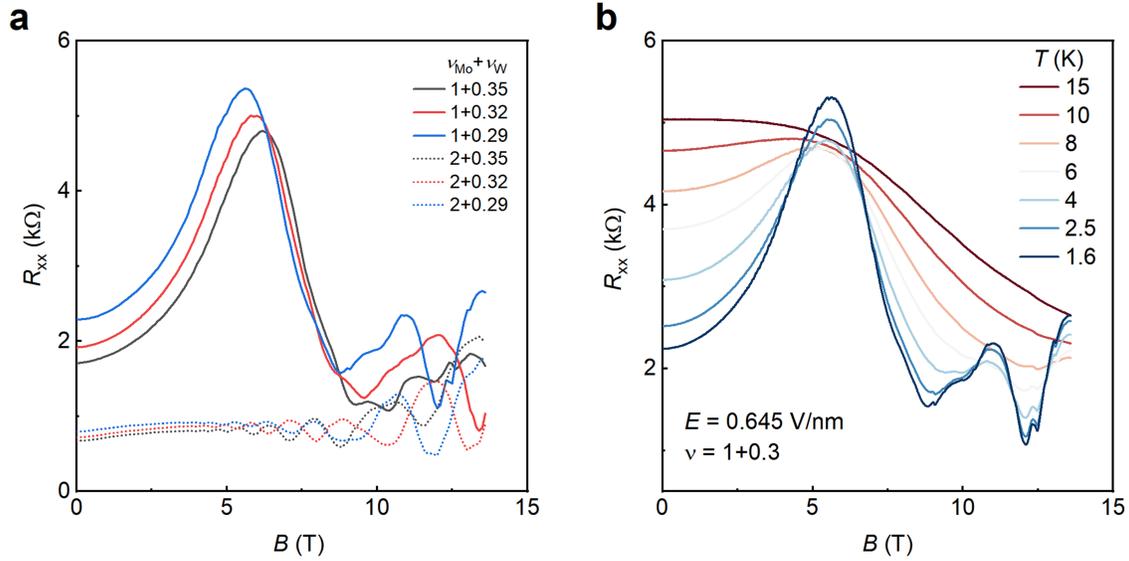

**Extended Data Figure 5 | Additional data on magnetic destruction of Kondo singlets. a**, Magnetic field dependence of $R_{xx}$ at $T = 1.6$ K in region II (solid curves) and III (dotted curves). Whereas a small magnetoresistance is observed in region III (except the quantum oscillations at high fields), a large, quadratic magnetoresistance is observed in region II before the magnetic destruction near 6 T. $R_{xx}$ for the two regions become comparable after the magnetic destruction. **b**, Magnetic field dependence of $R_{xx}$ at different temperatures in region II. The temperature dependence corresponds to that in Fig. 3f for $R_H$.

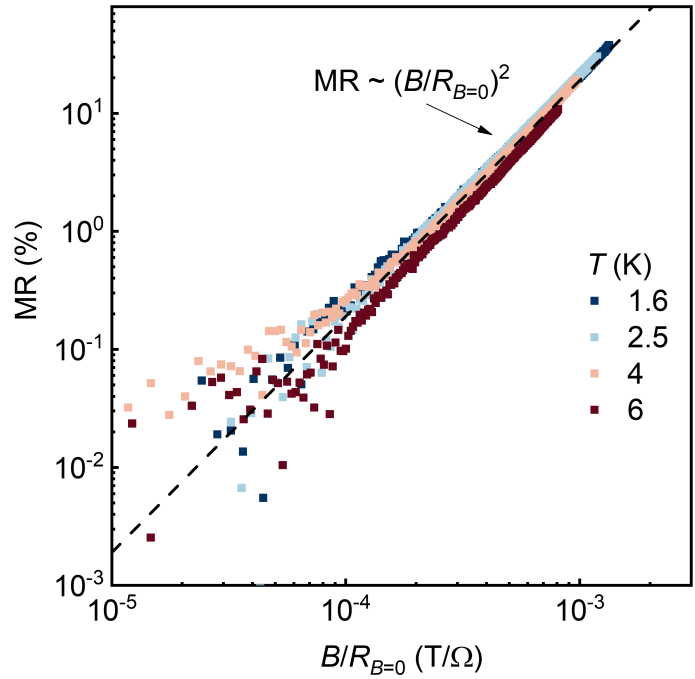

**Extended Data Figure 6 | Kohler's scaling in region II.** Magnetoresistance (MR) as a function of the scaled magnetic field (by the zero-field $R_{xx} \equiv R_{B=0}$) at $\nu = 1 + 0.3$ and varying temperatures. All curves collapse to the trend $MR \propto (\frac{B}{R_{B=0}})^2$ as shown by the black dashed line. The data further confirms the Fermi liquid behavior in region II.

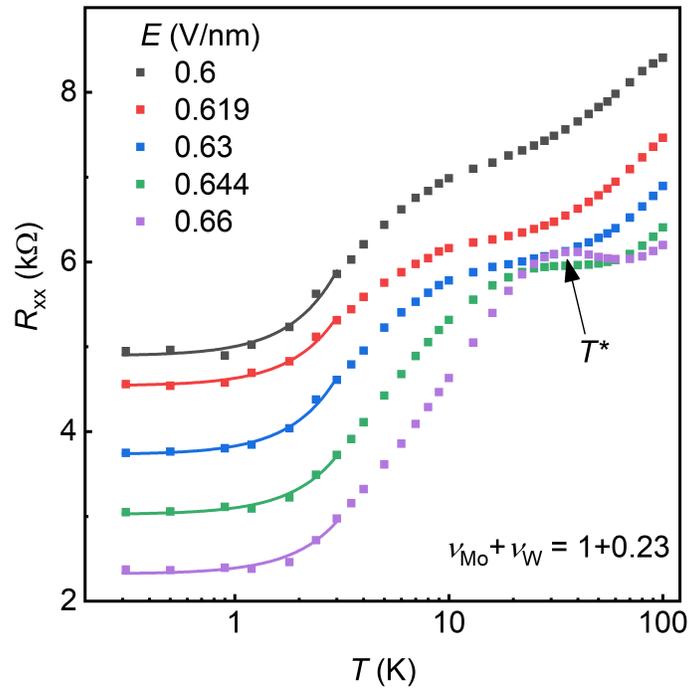

**Extended Data Figure 7 | Temperature dependence of $R_{xx}$ at varying electric field in region II.** The solid lines are the best fits to the quadratic temperature dependence at low temperatures. The resistance peak/bump determines the Kondo temperature $T^*$.

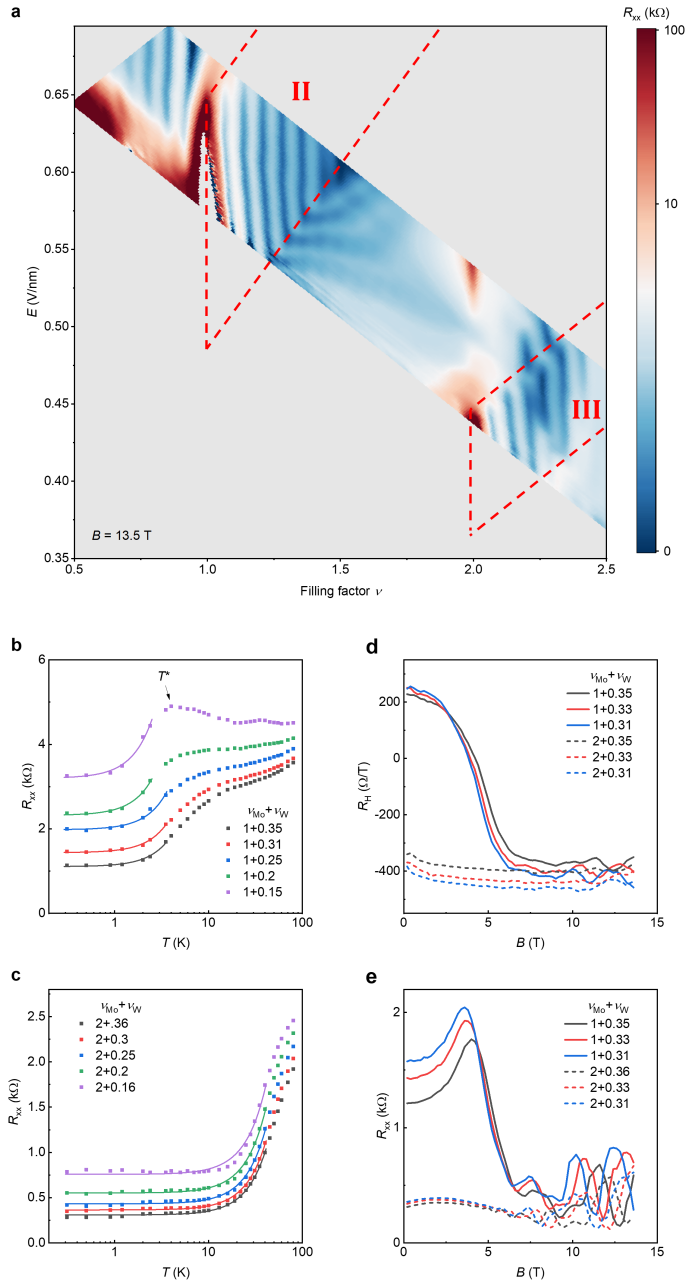

**Extended Data Figure 8 | Reproducible results in device 2. a,** Dependence of $R_{xx}$ on $\nu$ and $E$ at $B = 13.5$ T and $T = 1.6$ K. The red dashed lines mark the boundaries of region II and III. **b, c,** Temperature dependence of $R_{xx}$ at varying doping densities for $\nu = 1 + x$ (**b**) and $\nu = 2 + x$ (**c**). The solid lines are the best fits to the quadratic temperature dependence at low temperatures. **d, e,** Magnetic field dependence of $R_H$ (**d**) and $R_{xx}$ (**e**) at $T = 1.6$ K for both $\nu = 1 + x$ (solid curves) and $\nu = 2 + x$ (dashed curves). A sharp change in $R_H$ at the critical magnetic field is observed only for $\nu = 1 + x$; that at $\nu = 2 + x$ is nearly field independent.